\documentclass{PoS}

 \usepackage{graphicx,amsmath,amssymb}

\newcommand{\mbf}[1]{\mathbf{#1}}
\newcommand{\half}{{\frac{1}{2}}}

\title{Light-front holography - A new approach to  relativistic hadron dynamics and nonperturbative QCD}

\ShortTitle{Light-front holography}

\author{\speaker{Guy F. de T\'eramond}
\\
  Universidad de Costa Rica, San Jos\'e, Costa Rica\\
  E-mail: \email{gdt@asterix.crnet.cr}}

\author{Stanley J. Brodsky\\
\it SLAC National Accelerator Laboratory, \\
 Stanford University, Stanford, CA 94309, USA\\
        E-mail: \email{sjbth@slac.stanford.edu}
\thanks{This research was supported by the Department of Energy  contract DE--AC02--76SF00515. SLAC--PUB--15094.}}

\abstract{
The holographic mapping of gravity in AdS space to QCD, quantized at fixed light-front time, provides  a precise relation between the bound-state amplitudes in the fifth dimension of AdS space and the boost-invariant light-front wavefunctions describing the internal structure of hadrons in physical space-time.
In particular, the elastic and transition form factors of the pion and the nucleons are well described in this framework.  
The light-front AdS/QCD  holographic approach thus gives  a frame-independent first approximation of the color-confining dynamics,  spectroscopy, and excitation spectra of relativistic light-quark bound states in QCD.   
More generally, we show that the valence Fock-state wavefunctions of the eigensolutions of the light-front QCD Hamiltonian satisfy a single-variable relativistic equation of motion, analogous to the nonrelativistic radial Schr\"odinger equation, with an effective 
confining potential which systematically incorporates the effects of higher quark and gluon Fock states.  The proposed method to compute the effective interaction thus resembles the two-particle-irreducible functional techniques used in quantum field theory.
}

\FullConference{Sixth International Conference on Quarks and Nuclear Physics,\\
		April 16-20, 2012\\
		Ecole Polytechnique, Palaiseau, Paris}

\begin{document}

\section{Introduction}

Forty years after the discovery of QCD, the description of hadrons in terms of their fundamental quark and gluon constituents appearing in the QCD Lagrangian and the nature of color-confinement still remain  among the most challenging problems of strong interaction dynamics.
Euclidean lattice calculations provide an important numerical simulation of nonperturbative QCD. 
However, the excitation spectrum of hadrons represents an important challenge to lattice QCD due to the  enormous computational complexity beyond  ground-state configurations and the unavoidable presence of multi-hadron thresholds. In contrast, the incorporation of the AdS/CFT correspondence between gravity in AdS space and conformal field theories in physical space-time~\cite{Maldacena:1997re}  has led to an analytic  semiclassical approximation for strongly-coupled quantum field theories
as well as providing important new physical insight into the wavefunctions and nonperturbative dynamics of relativistic light-hadron bound states~\cite{deTeramond:2008ht}.

Light-front  (LF) holographic methods  were originally
introduced~\cite{Brodsky:2006uqa} by mapping  the electromagnetic form factors in AdS space~\cite{Polchinski:2002jw} to the corresponding  expression at fixed LF time in
physical space-time~\cite{Drell:1969km}.   It was also shown that one obtains  
an identical mapping for  the matrix elements of the energy-momentum tensor~\cite{Brodsky:2008pf}, by perturbing the AdS
metric around its static solution~\cite{Abidin:2008ku}.
In the ``bottom-up'' approach to  the gauge/gravity duality~\cite{Erlich:2005qh, DaRold:2005zs},  fields in the bulk geometry are introduced to match the
chiral symmetries of QCD. In contrast, in LF holography a direct connection with the internal constituent structure of hadrons is established using LF quantization
~\cite{deTeramond:2008ht, Brodsky:2006uqa, Brodsky:2008pf, Brodsky:2003px}.

The identification of  AdS space with partonic physics in physical space-time is specific to the light front: the transition amplitudes in AdS are expressed as a wavefunction overlap~\cite{Polchinski:2002jw} which maps precisely to the convolution of frame-independent LF wavefunctions (LFWFs)~\cite{Drell:1969km}.  In contrast, the AdS convolution formula cannot be mapped to current matrix elements at ordinary fixed time $t$  since one must include connected currents from the vacuum which are not given by eigensolutions of the instant-time Hamiltonian. There are no such vacuum  contributions in the LF  
for current matrix elements -- in agreement with the AdS formulae. Furthermore, the instant-time wavefunctions must be boosted 
from the hadron's rest frame -- an intractable dynamical problem.  

Unlike ordinary instant-time quantization, the Hamiltonian equation of motion in the LF is frame-independent and has a structure similar to  eigenmode equations in AdS space. This makes the direct connection of QCD to AdS/CFT methods possible.  In fact, one can also study the AdS/CFT duality and its modifications starting from the LF Hamiltonian equation of motion for a relativistic bound-state system in physical space-time~\cite{deTeramond:2008ht}. To a first semiclassical approximation, where quantum loops and quark masses are not included, LF holography leads to a LF Hamiltonian equation which
describes the bound-state dynamics of light hadrons in terms of an invariant impact kinematical variable $\zeta$ which measures the
separation of the partons within the hadron at equal LF time. Remarkably, the unmodified AdS equations
correspond to the kinetic energy terms of  the partons inside a
hadron, whereas the interaction terms in the QCD Lagrangian build confinement and
correspond to the truncation of AdS space in an effective dual
gravity  approximation~\cite{deTeramond:2008ht}.  Thus, all the complexities of strong-interaction dynamics are hidden in an effective 
confining potential $U(\zeta)$, which acts in the valence sector of the theory, reducing the many-particle problem in QCD to an effective one-body problem. The derivation of the effective interaction  $U(\zeta)$ directly from QCD then becomes the central issue.

\section{The Light-front Schr\"odinger equation: a semiclassical approximation to QCD \label{LFQCD}}

The hadronic four-momentum  generator in the front form~\cite{Dirac:1949cp} is denoted by $P =  (P^+, P^-, \mbf{P}_{\!\perp})$, where  the longitudinal and transverse generators  $P^+$ and  $\mbf{P}_\perp$ do not depend on the interaction (they are  kinematical generators which leave the LF plane invariant) and the dynamical generator $P^-$ which contain the interactions. It is the LF time $x^+= x^0 + x^3$ evolution operator, $i \frac{\partial}{\partial x^+} \vert \psi \rangle = P^- \vert \psi\rangle$, and it is constructed canonically from the QCD Lagrangian~\cite{Brodsky:1997de}.  The
hadronic mass states $P_\mu P^\mu = M^2$ are determined by the
Lorentz-invariant Hamiltonian equation for the relativistic bound-state 
\begin{equation} \label{LFH}
H_{LF} \vert  \psi(P) \rangle =  M^2 \vert  \psi(P) \rangle,
\end{equation}
with  $H_{LF} \equiv P_\mu P^\mu  =  P^- P^+ -  \mbf{P}_\perp^2$. The hadronic state $\vert\psi\rangle$ is an expansion in multiparticle Fock states
$\vert \psi \rangle = \sum_n \psi_n \vert n \rangle$, where the components $\psi_n = \langle n \vert \psi \rangle$ are a column vector of states, and the basis vectors $\vert n \rangle$ are the $n$-parton eigenstates of the free LF Hamiltonian: 
$\vert q \bar q \rangle , \vert q \bar q g \rangle,   \vert q \bar q  q \bar q \rangle  \cdots$ etc.

For certain applications it is useful to reduce the multiparticle eigenvalue problem (\ref{LFH}) to a single equation~\cite{Pauli:1998tf}, instead of diagonalizing the Hamiltonian. The central problem then becomes the derivation of the effective interaction, which acts only on the valence sector of the theory and has, by definition, the same eigenvalue spectrum as the initial Hamiltonian problem. For carrying out this program one most systematically express the higher Fock components as functionals of the lower ones. The method has the advantage that the Fock space is not truncated and the symmetries of the Lagrangian are preserved~\cite{Pauli:1998tf}.

In our recent work we have shown how light front holographic methods lead to a remarkably simple equation of motion for mesons at fixed light-front time.  To this end, we write the LFWF in terms of the invariant impact-space variable for a two-parton state $\zeta^2= x(1-x)\mbf{b}_\perp^2$ 
\begin{equation} \label{eq:psiphi}
\psi(x,\zeta, \varphi) = e^{i L \varphi} X(x) \frac{\phi(\zeta)}{\sqrt{2 \pi \zeta}} ,
\end{equation}
thus factoring the angular dependence $\varphi$ and the longitudinal, $X(x)$, and transverse mode $\phi(\zeta)$.
In the limit of zero quark masses the longitudinal mode decouples and
the LF eigenvalue equation $P_\mu P^\mu \vert \phi \rangle  =  M^2 \vert \phi \rangle$
is thus a light-front  wave equation for $\phi$
\begin{equation} \label{LFWE}
\left(-\frac{d^2}{d\zeta^2}
- \frac{1 - 4L^2}{4\zeta^2} + U\left(\zeta^2, J\right) \right)
\phi_{J,L,n}(\zeta^2) = M^2 \phi_{J,L,n}(\zeta^2),
\end{equation}
a relativistic single-variable  LF  Schr\"odinger equation (LFSE). The effective interaction $U$ is instantaneous in LF time and acts on the lowest state of the LF Hamiltonian.  This equation describes the spectrum of mesons as a function of $n$, the number of nodes in $\zeta^2$, the total angular momentum $J=J^z$ and the internal orbital angular momentum of the constituents $L= L^z$~\footnote{The  $SO(2)$ Casimir  $L^2$  corresponds to  the group of rotations in the transverse LF plane.}. 
It is the relativistic frame-independent front-form analog of the non-relativistic radial Schr\"odinger equation for muonium  and other hydrogenic atoms in presence of an instantaneous Coulomb potential.

\section{Effective confinement interaction from the gauge/gravity correspondence}

A remarkable correspondence between the equations of motion in AdS  and the Hamiltonian equation for relativistic bound-states  was found in~\cite {deTeramond:2008ht}.   In fact, to a first semiclassical approximation,
LF QCD  is formally equivalent to the equations of motion on a fixed gravitational background~\cite{deTeramond:2008ht} asymptotic to AdS$_5$, where  confinement properties  are encoded in a dilaton profile $\varphi(z)$.

A spin-$J$ field in AdS$_{d+1}$ is represented by a rank $J$ tensor field $\Phi_{M_1 \cdots M_J}$, which is totally symmetric in all its indices.  In presence of a dilaton background field $\varphi(z)$ the action is~\footnote{The study of   higher integer and half-integer spin wave equations  in  AdS  is based on our collaboration with Hans Guenter Dosch.  See also the discussion in Ref.~\cite{Gutsche:2011vb}.}
\begin{multline} \label{SJ}
S = \half \int \! d^d x \, dz  \,\sqrt{g} \,e^{\varphi(z)}
  \Big( g^{N N'} g^{M_1 M'_1} \cdots g^{M_J M'_J} D_N \Phi_{M_1 \cdots M_J} D_{N'}  \Phi_{M'_1 \cdots M'_J}    \\
 - \mu^2  g^{M_1 M'_1} \cdots g^{M_J M'_J} \Phi_{M_1 \cdots M_J} \Phi_{M'_1 \cdots M'_J}  + \cdots \Big)  ,
\end{multline}
where $M, N = 1, \cdots , d+1$, $\sqrt{g} = (R/z)^{d+1}$ and $D_M$ is the covariant derivative which includes parallel transport. 
The coordinates of AdS are the Minkowski coordinates $x^\mu$ and the holographic variable $z$ labeled $x^M = \left(x^\mu, z\right)$. The  d + 1 dimensional mass $\mu$ is not a physical observable and is {\it a priory} an arbitrary
parameter. The dilaton background field $\varphi(z)$ in  (\ref{SJ})   introduces an energy scale in the five-dimensional AdS action, thus breaking its conformal invariance. It  vanishes in the conformal ultraviolet limit $z \to 0$.

 A physical hadron has plane-wave solutions and polarization indices along the 3 + 1 physical coordinates
 $\Phi_P(x,z)_{\mu_1 \cdots \mu_J} = e^{- i P \cdot x} \Phi(z)_{\mu_1 \cdots \mu_J}$,
 with four-momentum $P_\mu$ and  invariant hadronic mass  $P_\mu P^\mu \! = M^2$. All other components vanish identically. 
 One can then construct an effective action in terms
 of the spin modes $\Phi_J = \Phi_{\mu_1 \mu_2 \cdots \mu_J}$ with only  physical degrees of 
 freedom. In this case the system of coupled differential equations which follow from (\ref{SJ}) reduce to a homogeneous equation in terms of the physical field $\Phi_J$
  upon rescaling the AdS mass $\mu$
 \begin{equation} \label{AdSWEJ}
\left[-\frac{ z^{d-1 -2 J}}{e^{\varphi(z)}}   \partial_z \left(\frac{e^\varphi(z)}{z^{d-1 - 2 J}} \partial_z\right)
+ \left(\frac{\mu R}{z}\right)^2\right] \Phi(z)_J = M^2 \Phi(z)_J .
 \end{equation}

 Upon the substitution $z \! \to\! \zeta$  and 
$\phi_J(\zeta)   = \left(\zeta/R\right)^{-3/2 + J} e^{\varphi(z)/2} \, \Phi_J(\zeta)$ 
in (\ref{AdSWEJ}), we find for $d=4$ the LFSE (\ref{LFWE}) with effective potential~\cite{deTeramond:2010ge}
\begin{equation} \label{U}
U(\zeta^2, J) = \half \varphi''(\zeta^2) +\frac{1}{4} \varphi'(\zeta^2)^2  + \frac{2J - 3}{2 \zeta} \varphi'(\zeta^2) ,
\end{equation}
provided that the fifth dimensional mass $\mu$ is related to the internal orbital angular momentum $L = max \vert L^z \vert$ and the total angular momentum $J^z = L^z + S^z$ according to $(\mu R)^2 = - (2-J)^2 + L^2$.  The critical value  $L=0$  corresponds to the lowest possible stable solution, the ground state of the LF Hamiltonian.
For $J = 0$ the five dimensional mass $\mu$
 is related to the orbital  momentum of the hadronic bound state by
 $(\mu R)^2 = - 4 + L^2$ and thus  $(\mu R)^2 \ge - 4$. The quantum mechanical stability condition $L^2 \ge 0$ is thus equivalent to the Breitenlohner-Freedman stability bound in AdS~\cite{Breitenlohner:1982jf}.

The correspondence between the LF and AdS equations  thus determines the effective confining interaction $U$ in terms of the infrared behavior of AdS space and gives the holographic variable $z$ a kinematical interpretation. The identification of the orbital angular momentum 
is also a key element of our description of the internal structure of hadrons using holographic principles.

A particularly interesting example is a dilaton profile $\exp{\left(\pm \kappa^2 z^2\right)}$ of either sign, since it 
leads to linear Regge trajectories~\cite{Karch:2006pv} and avoids the ambiguities in the choice of boundary conditions at the infrared wall.  
For the  confining solution $\varphi = \exp{\left(\kappa^2 z^2\right)}$ the effective potential is
$U(\zeta^2,J) =   \kappa^4 \zeta^2 + 2 \kappa^2(J - 1)$ and  Eq.  (\ref{LFWE}) has eigenvalues
$M_{n, J, L}^2 = 4 \kappa^2 \left(n + \frac{J+L}{2} \right)$,
with a string Regge form $M^2 \sim n + L$.  
A discussion of the light meson and baryon spectrum,  as well as  the elastic and transition form factors of the light hadrons using LF holographic methods, is given in Ref.~\cite{deTeramond:2012rt}.

\section{Effective confinement interaction from higher Fock states in light-front QCD}

As we have discussed in Sec. \ref{LFQCD}, one can systematically eliminate the higher Fock states in terms of an effective interaction $U(\zeta^2, J ,M^2)$  in order to obtain an equation for the valence $\vert q \bar q \rangle$ Fock state~\cite{Pauli:1998tf}. 
The potential $U$ depends on the eigenvalue $M^2$ via the LF energy denominators 
$P^-_{\rm initial} - P^-_{\rm intermediate} + i \epsilon$
of the intermediate states which connect different LF Fock states.  
Here $P^-_{\rm initial} =( {M^2 + \mathbf{P}^2_\perp)/ P^+}$. The dependence of $U$ on $M^2$ is analogous to the retardation effect in QED interactions, such as the hyperfine splitting in muonium, which involves the exchange of a propagating photon.  Accordingly, the eigenvalues $M^2$ must be determined self-consistently.
The  $M^2$ dependence of the effective potential thus reflects the contributions from higher Fock states in the LFSE (\ref{LFWE}),  since
 $U(\zeta^2, J ,M^2)$ is also the kernel for the  scattering amplitude $q \bar q \to q \bar q$ at $s = M^2.$    It has only ``proper'' contributions; {\it i.e.}, it has no $q \bar q$ intermediate state.  The potential can be constructed systematically using LF time-ordered perturbation theory.  Thus the QCD theory has identical form as the AdS theory, but with the quantum field-theoretic corrections due to the higher Fock states giving a general form for the potential.  This provides a novel way to solve nonperturbative QCD.

This LFSE for QCD becomes increasingly accurate as one includes contributions from very high particle number Fock states. There is only one dynamical variable $\zeta^2$.  The AdS/QCD harmonic oscillator potential could emerge when one includes contributions from the exchange of two connected gluons; {\it i.e.}, ``H'' diagrams~\cite{Appelquist:1977tw}.
We notice that $U$ becomes complex for an excited state since a denominator can vanish; this gives a complex eigenvalue and the decay width.

The above discussion assumes massless quarks. More generally we must include mass terms
${m^2_a / x}  + {m^2_b/(1-x)}$ in the kinetic energy term and allow  the potential $  U(\zeta^2, x, J,M^2)$ to have dependence on the LF momentum fraction $x$.  The quark masses also appear in $U$ due to the  presence in the LF denominators as well as the chirality-violating interactions connecting the valence Fock state to the higher Fock states. In this case, however, the equation of motion cannot be reduced to a single variable.

The LFSE approach also can be applied to atomic bound states in QED and nuclei. In principle one could compute the spectrum and dynamics of atoms, such as the Lamb shift and hyperfine splitting of hydrogenic atoms to high precision by a systematic treatment of the potential. Unlike the ordinary  instant form, the resulting LFWFs are independent of the total momentum and can thus describe ``flying atoms'' without the need for dynamical boosts, 
such as the ``true muonium'' $(\mu^+ \mu^-)$ bound states which can be produced  by Bethe-Heitler pair production $\gamma \, Z \to (\mu^+ \mu^-)  \, Z$  below threshold~\cite{Brodsky:2009gx}. A related approach for determining the valence light-front wavefunction and studying the effects of higher Fock states without truncation has been given in Ref.~\cite{Chabysheva:2011ed}.

\section{Conclusions}

Despite some limitations, AdS/QCD, the LF holographic  approach to the gauge/gravity duality,  has  given significant physical  insight into the strongly-coupled nature and internal structure of hadrons.  In particular, the AdS/QCD soft-wall model provides an elegant analytic framework for describing  nonperturbative  hadron dynamics, the systematics of the excitation spectrum of hadrons, including their empirical multiplicities and degeneracies. It also provides powerful new analytical tools for computing hadronic transition amplitudes incorporating conformal scaling behavior at short distances and the transition from the hard-scattering perturbative domain, where quark and gluons are the relevant degrees of freedom, to the long-range confining hadronic region.   We have also discussed the possibility of  computing the effective confining potential in light-front QCD for  a single-variable LF Schr\"odinger equation by systematically incorporating the effects of higher  Fock states, thus providing  the basis for a profound connection between physical QCD, quantized on the light-front, and the physics of hadronic modes in a higher dimensional AdS space.


\begin{thebibliography}{99}


  
  \bibitem{Maldacena:1997re}
 J.~M.~Maldacena,
 \href{http://www.springerlink.com/content/q508214382421612/}{  Int.\ J.\ Theor.\ Phys.\  {\bf 38}, 1113 (1999)}.
  
\bibitem{deTeramond:2008ht}
  G.~F.~de Teramond and S.~J.~Brodsky,
  \href{http://prl.aps.org/abstract/PRL/v102/i8/e081601}{ Phys.\ Rev.\ Lett.\  {\bf 102}, 081601 (2009)}.
 
 
  \bibitem{Brodsky:2006uqa}
  S.~J.~Brodsky and G.~F.~de Teramond,
  \href{http://prl.aps.org/abstract/PRL/v96/i20/e201601}{ Phys.\ Rev.\ Lett.\  {\bf 96}, 201601 (2006)};
  \href{http://prd.aps.org/abstract/PRD/v77/i5/e056007}{ Phys.\ Rev.\  D {\bf 77}, 056007 (2008)}.
  
  
  \bibitem{Polchinski:2002jw}
  J.~Polchinski and M.~J.~Strassler,
  \href{http://iopscience.iop.org/1126-6708/2003/05/012/}{ HEP {\bf 0305}, 012 (2003)}.
  
  
  \bibitem{Drell:1969km}
  S.~D.~Drell and T.~M.~Yan,
  \href{http://prl.aps.org/abstract/PRL/v24/i4/p181_1}{ Phys.\ Rev.\ Lett.\  {\bf 24}, 181 (1970)};
  G.~B.~West,
  \href{http://prl.aps.org/abstract/PRL/v24/i21/p1206_1}{ Phys.\ Rev.\ Lett.\  {\bf 24}, 1206 (1970)}.
  
  
  \bibitem{Brodsky:2008pf}
 S.~J.~Brodsky and G.~F.~de Teramond,
  \href{http://prd.aps.org/abstract/PRD/v78/i2/e025032}{ Phys.\ Rev.\  D {\bf 78}, 025032 (2008)}.
  
  
   \bibitem{Abidin:2008ku}
  Z.~Abidin and C.~E.~Carlson,
  \href{http://prd.aps.org/abstract/PRD/v77/i9/e095007}{ Phys.\ Rev.\  D {\bf 77}, 095007 (2008)}.
  
  
   \bibitem{Erlich:2005qh}
 J.~Erlich, E.~Katz, D.~T.~Son and M.~A.~Stephanov,
  \href{http://prl.aps.org/abstract/PRL/v95/i26/e261602}{Phys.\ Rev.\ Lett.\  {\bf 95}, 261602 (2005)}.

 \bibitem{DaRold:2005zs}
  L.~Da Rold and A.~Pomarol,
 \href{http://www.sciencedirect.com/science/article/pii/S0550321305004074}{Nucl.\ Phys.\ B {\bf 721}, 79 (2005)}.
  
   \bibitem{Brodsky:2003px} 
  S.~J.~Brodsky and G.~F.~de Teramond,
  \href{http://www.sciencedirect.com/science/article/pii/S037026930400053X}{Phys.\ Lett.\ B {\bf 582}, 211 (2004)}.
 
 \bibitem{Dirac:1949cp}
  P.~A.~M.~Dirac,
 \href{http://rmp.aps.org/abstract/RMP/v21/i3/p392_1}{ Rev.\ Mod.\ Phys.\  {\bf 21}, 392 (1949)}.
 
  \bibitem{Brodsky:1997de}
  S.~J.~Brodsky, H.~C.~Pauli and S.~S.~Pinsky,
  \href{http://www.sciencedirect.com/science/article/pii/S0370157397000896}{Phys.\ Rept.\  {\bf 301}, 299 (1998)}.
 
\bibitem{Pauli:1998tf} 
  H.~C.~Pauli,
  \href{http://www.springerlink.com/content/yxkbbv68lmk9n7lc/?MUD=MP}{Eur.\ Phys.\ J.\ C {\bf 7}, 289 (1999)}.
  
  \bibitem{Gutsche:2011vb}
  T.~Gutsche, V.~E.~Lyubovitskij, I.~Schmidt, A.~Vega,
\href{http://prd.aps.org/abstract/PRD/v85/i7/e076003}{Phys.\ Rev.\ D {\bf 85}, 076003 (2012)}.
   
  
   \bibitem{deTeramond:2010ge}
  G.~F.~de Teramond and S.~J.~Brodsky,
\href{http://scitation.aip.org/getabs/servlet/GetabsServlet?prog=normal&id=APCPCS001296000001000128000001&idtype=cvips&gifs=yes&ref=no}
{AIP Conf.\ Proc.\  {\bf 1296}, 128 (2010)}.
  
  \bibitem{Breitenlohner:1982jf}
  P.~Breitenlohner and D.~Z.~Freedman,
  \href{http://www.sciencedirect.com/science/article/pii/0003491682901166}{Annals Phys.\  {\bf 144}, 249 (1982)}.
  
  \bibitem{Karch:2006pv}
 A.~Karch, E.~Katz, D.~T.~Son and M.~A.~Stephanov,
 \href{http://prd.aps.org/abstract/PRD/v74/i1/e015005}{Phys.\ Rev.\  D {\bf 74}, 015005 (2006)}.
 
 \bibitem{deTeramond:2012rt} 
 G.~F.~de Teramond and S.~J.~Brodsky,
  \href{http://arXiv.org/abs/arXiv:1203.4025}{arXiv:1203.4025 [hep-ph]}.
  
  \bibitem{Appelquist:1977tw} 
  T.~Appelquist, M.~Dine and I.~J.~Muzinich,
  \href{http://www.sciencedirect.com/science/article/pii/0370269377906517}{Phys.\ Lett.\ B {\bf 69}, 231 (1977)}.
 
 \bibitem{Brodsky:2009gx}
S.~J.~Brodsky and R.~F.~Lebed,
\href{http://prl.aps.org/abstract/PRL/v102/i21/e213401}{Phys.\ Rev.\ Lett.\  {\bf 102}, 213401 (2009)}.

\bibitem{Chabysheva:2011ed} 
 S.~S.~Chabysheva and J.~R.~Hiller,
\href{http://www.sciencedirect.com/science/article/pii/S0370269312004352}{Phys.\ Lett.\ B {\bf 711}, 417 (2012).}

\end{thebibliography}
\end{document}